\documentclass[12pt,a4paper,oneside]{article}

\usepackage{graphicx}

\usepackage{subfigure}

\usepackage[T2C,OT1]{fontenc}
\usepackage[utf8]{inputenc}
\usepackage[russian,english]{babel}

\usepackage[svgnames]{xcolor}

\usepackage{sectsty}

\usepackage{fancyhdr}

\usepackage[breaklinks,linktoc=page]{hyperref}

\sectionfont{\normalsize \centering}
\subsectionfont{\normalsize \centering}
\begin{document}

\begin{center}
  {\large\bf Observation of structures at $\sim 17$ and $\sim 38$ MeV/c$^{2}$ in the\\
   $\gamma\gamma$ invariant mass spectra in pC, dC, and dCu collisions\\
    at $\textit{p}_{lab}$ of a few GeV/c per nucleon\\}
\vspace{0.40cm}

{\bf Kh.U. Abraamyan$^{1,2 *}$, Ch. Austin$^{3}$, M.I.
Baznat$^{4}$,
K.K.~Gudima$^{4}$,
M.A. Kozhin$^{1}$,
S.G. Reznikov$^{1}$, and A.S. Sorin$^{1,5}$}\\
\vspace{0.40cm}

{$^1$VBLHEP JINR, 141980 Dubna, Moscow region, Russia\\
$^2$International Center for Advanced Studies, YSU, 0025, Yerevan, Armenia\\
$^3$ 33 Collins Terrace, Maryport, Cumbria CA15 8DL, England\\
$^4$ Institute of Applied Physics, MD-2028 Kishinev, Moldova\\
$^5$ BLTP JINR, 141980 Dubna, Moscow region, Russia\\}
\vspace{0.40cm}

% {* E-mail: abraam@sunhe.jinr.ru}
{* E-mail: \href{mailto:abraamyan@jinr.ru}{abraamyan@jinr.ru}}
\end{center}
\vspace{0.10cm}

\noindent The results of an analysis of the invariant mass spectra of photon
pairs produced in $d$C, $p$C and $d$Cu interactions at momenta of 2.75, 5.5 and 3.83 GeV/$c$ per nucleon respectively, are presented. Signals
in the form of enhanced structures at invariant masses of about 17 and 38 MeV/$c^2$ are observed.
The results of testing of the observed signals, including the results of the
Monte Carlo simulation are presented. The test results support the conclusion
that the observed signals are the consequence of detection of the particles with
masses of about 17 and 38 MeV/$c^2$ decaying into a pair of photons.
\vspace{0.40cm}

%\begin{otherlanguage*}{russian}
{\selectlanguage{russian}
\noindent Представлены результаты анализа спектров инвариантных масс пар
фотонов, образуемых в $d$C-, $p$C- и $d$Cu-взаимодействиях при импульсах 2,75, 5,5 и 3,83 ГэВ/$c$ на нуклон соответственно. Наблюдаются превышения в виде структур при инвариантных массах около 17 и 38 МэВ/$c^2$. Приведены результаты проверки наблюдаемых сигналов, в том числе
результаты моделирования по методу Монте-Карло. Результаты проверки подтверждают
вывод о том, что \mbox{наблюдаемые} сигналы являются следствием регистрации
частиц с массами около 17 и 38 МэВ/$c^2$, \mbox{распадающихся} на пару фотонов.
%\end{otherlanguage*}
}

\vspace{0.15cm}

\section{INTRODUCTION}
\label{sec: INTRODUCTION}

A series of experiments on the production of photon pairs in the interactions of protons, deuterons and alpha particles with nuclei was carried out on the internal beams of the Nuclotron at JINR. The experiments were performed on a multichannel two-arm gamma spectrometer of the SPHERE setup (the PHOTON-2 setup). The results of the first analysis on the production of $\eta$ mesons (selection of photons from different arms of the spectrometer) have been published in
\cite{R1}.

At the suggestion of E. van Beveren and G. Rupp \cite{van Beveren Rupp First indications},
the spectra of photon pairs in the region of invariant masses around
38 MeV/c2 were analyzed in order to search for the E38 boson.
The results of this analysis (photons from the same spectrometer arm)
are published in \cite{Ab2019}.

In recent experiments in the Institute for Nuclear Research (ATOMKI) \cite{Atomki3}, an anomalous correlation between the opening angles and the total energies of
 $e^+e^-$ pairs was observed at the invariant mass of the pairs of about 17 MeV/$c^2$, which can be interpreted as the result of production and decay of a light boson, called the $X17$ particle.

This anomaly is currently being widely discussed \cite{Rome2021}.
Various models are proposed that attempt to describe the observed anomaly at 17 MeV/$c^2$:
the search for new physics (the fifth-force interpretation) \cite{Fifth};
an axion \cite{axion};
resonant production mechanism \cite{ResMec};
calculations in frame of effective field theories \cite{EffF};
a model for different EM transitions and interferences \cite{EMinterf};
calculations of particle masses in the open-string model in
 two-dimensional quantum chromodynamics and quantum electrodynamics model
\cite{Wong1, Wong2, Wong3Fig3} and in the flux tube model \cite{Dyach};
an attempt to find AU(1)' solution of the 17 MeV anomaly \cite{AU1}.
In particular, in \cite{Wong3Fig3} and in an earlier work \cite{Wong2}, it is proposed that a light quark and a light antiquark may be bound and confined by the QED interaction as a neutral isoscalar boson at 17 and a neutral isovector boson at 38 MeV, with the QED $q\widetilde{q}$ isoscalar composite as a possible candidate for the X17 boson.

Besides this, in recent years, the possibility of the existence of a light scalar meson -
lighter than the known lightest hadron – $\pi^0$-meson, has been actively
discussed \cite{Ab2019}. As noted in \cite{Wolkanowski}, the issue of scalar mesons has been
the subject of vivid debate among the physical community for a long time because
their identification and explanation in terms of quarks and gluons is difficult,
see e.g.\hspace{-0.7ex} Refs.\hspace{-0.7ex} \cite{Amsler, Amsler Close,
Parganlija Giacosa, Parganlija et al} and references therein.

A first indication of the existence of a 38 MeV light scalar boson
(henceforth referred to as the $E(38)$) was reported in \cite{van Beveren Rupp First
indications}, where some evidence for the existence of a light scalar particle
that most probably couples exclusively to gluons and quarks is presented.
Theoretical and phenomenological arguments are presented to support the
existence of a light scalar boson for confinement and quark-pair creation.
Previously observed interference effects allow to set a narrow window for the
scalar's mass and also for its flavor-mass-dependent coupling to quarks.

Using data from a preliminary report \cite{Abraamyan et al Aug 12} on the
measurements presented in the present paper, and a simple model for the
collision of a deuteron or proton with a target nucleus, a very rough estimate
of the coupling constant of the $E(38)$ boson to the light quarks was obtained in
\cite{Austin}, assuming that the $E(38)$ boson is produced in a
bremsstrahlung-like manner and decays only to two photons.

From the latest results, in particular, in \cite{Aoki et al}, based on lattice
simulations using highly improved staggered quarks for twelve-flavor QCD with
several bare fermion masses, a flavor-singlet scalar state {\bf lighter than the
pion} in the correlators of fermionic interpolating operators is observed. The
same state is also investigated using correlators of gluonic interpolating
operators. Combined with their previous study, that showed twelve-flavor QCD to
be consistent with being in the conformal window, the authors of \cite{Aoki et
al} infer that the lightness of the scalar state is due to infrared
conformality. This result shed some light on the possibility of a light
composite Higgs boson (``technidilaton'') in walking technicolor theories.

In view of the above many possibilities, it is of great to search for possible particles in this region.
A very good method to produce these anomalous particles is by relativistic nucleus-nucleus collisions,
including proton collisions because the anomalous particles will likely involve quarks and antiquarks.
The search effort can be readily facilitated by studying the diphoton decay products of such particles,
as it has been demonstrated with our previous work and apparatus on the successful production and detection
of $\pi^0$ and $\eta$ mesons. For this anomalous region, it is important to confirm the observation of the
$X17$ particle using very different techniques and apparatus. It is furthermore important to know whether
the $X17$, $E38$, and $\pi^0$ can be observed by the same set of apparatus.
The simultaneous observation of the all three particles, places a severe constraint on the construction of
models for the anomalous particles and the nature of this region.

In our experiments, we measured both the energies and the coordinates of the photons
and thus measured the invariant mass of photon pairs.
The collected statistics made it possible to obtain, after  the background subtraction,
statistically significant signals in the range of invariant masses both about 17 and about 38 MeV/$c^2$.

\section{EXPERIMENT}
\label{sec: EXPERIMENT}
\subsection{General layout}
\label{sec: General layout}

The data acquisition of production of neutral mesons and $\gamma$-quanta in
interactions of protons and light nuclei with nuclei has been carried out with
internal beams of the JINR Nuclotron \cite{R1}. The experiments were conducted
with internal proton beams at momentum 5.5 GeV/$c$ incident on a carbon target
and with $^2$H, $^4$He beams and internal C-, Al-, Cu-, W-, Au-targets at
momenta from 1.7 to 3.8 GeV/$c$ per nucleon. For the first analysis the data for
the $d(2.0 A\textrm{ GeV}) + \mathrm{C}$, $d(3.0 A\textrm{ GeV}) + \mathrm{Cu}$ and
$p(4.6\textrm{ GeV}) + \mathrm{C}$ reactions were selected. Some results
on $\gamma \gamma$ pair production in these reactions, for the effective mass region, $M_{\gamma \gamma} > \textrm{100 MeV}/c^2$
(photons in pair from different arms of the spectrometer) were reported in
\cite{R1}.

Typical proton and deuteron fluxes were about $10^8$ and $10^9$ per pulse respectively.
The electromagnetic lead glass calorimeter PHOTON-2 was used to measure both the
energies and emission angles of photons. The experimental
instrumentation is schematically represented in Fig. \ref{Experimental
instrumentation}.

% ExperimentalSetup.png
\begin{figure}[htb]
\centering
\includegraphics[width=0.654\textwidth]{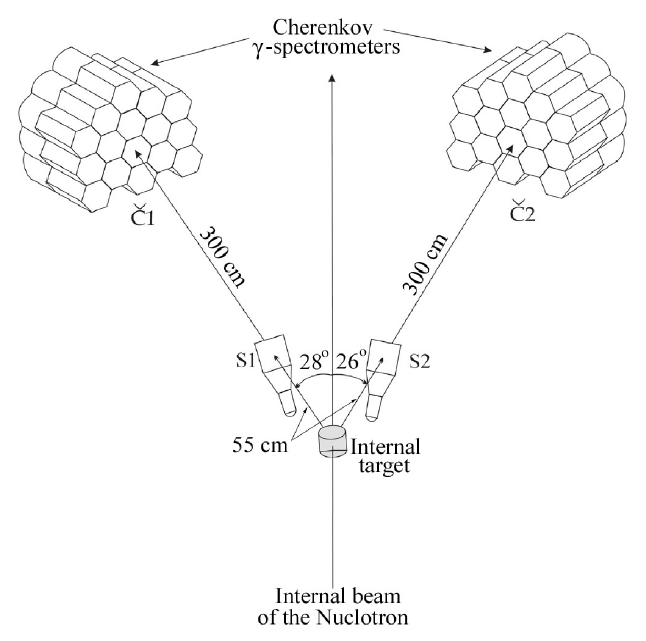}
\caption{The schematic drawing of the experimental PHOTON-2 setup. The S1 and S2
are scintillation counters.}
\label{Experimental instrumentation}
\end{figure}

The PHOTON-2 setup includes 32 $\gamma$-spectrometers of lead glass and
scintillation counters S1 and S2 of $2 \times 15 \times 15$ cm$^3$ used in front
of the lead glass for the charged particle detection \cite{Kha85}.

The center of the front surfaces of the lead glass hodoscopes is located at 300
cm from the target and at angles 25.6$^\circ$ and 28.5$^\circ$ with respect to
the beam direction. This gives a solid angle of 0.094 sr (0.047 sr for each
arm). Some details of the construction and performance of the lead glass
hodoscope are given in Tabl.\hspace{-0.5ex} \ref{Hodoscope parameters}.  The
internal target consists of carbon wires with the diameter of 8 microns, or
 a copper wire with the diameter of 50 microns mounted in a
rotatable frame. The overall construction is located in the vacuum shell of the
accelerator.

% Uncomment the following \begin{table} line instead of the one after it, to
% put Table 1 on page 3, under Fig 1
% \begin{table}[!htb]
\begin{table}[htb]
 \begin{tabular}{|l|c|}
\hline
Number of lead glasses & 32 TF-1, total weight 1422 kg \\
Module cross section & r = 9 cm of insert circumference \\
Module length & 35 cm, 14 R.L. \\
Spatial resolution & 3.2 cm \\
Angular resolution & 0.6$^\circ$ \\
Energy resolution & (3.9 / $\sqrt{E}$ + 0.4)\%, $E$ [GeV] \\
Gain stability & (1-2)\% \\
Dynamic range & 50 MeV - 6 GeV \\
Minimum ionizing signal & 382 $\pm$ 4 MeV of the photon equivalent \\
Total area & 0.848 m$^2$ \\
\hline

 \end{tabular}
\caption{\label{Hodoscope parameters}The basic parameters of the lead
glass hodoscope.}
\end{table}

Before the experiment the energy calibration of the lead glass counters has been
carried out  with 1.5 GeV/c per nucleon deuteron-beam at the JINR
synchrophasotron \cite{Ab89}. The long-term gain stability was continuously
monitored for each of the lead glass modules with 32 NaI(Tl) crystals doped with
$^{241}$Am sources.
Amplitude spectra from these sources in separate modules and spectra in the same modules in the experiment are given in Appendix 1.

The modules of the $\gamma$-spectrometer are assembled into two arms of 16
units. The modules in each arm are divided into two groups of 8 units. The
output signals of each group from 8 counters are summed up linearly and sent
to the inputs of four discriminators ($D_i$).
In these experiments the discriminator thresholds
were at the level of 0.4 GeV
for the $p + \mathrm{C}$ and $d + \mathrm{C}$ reactions
and 0.35 GeV for the $d + \mathrm{Cu}$ reaction.
Triggering takes place when there is a coincidence of signals from two or more groups from different arms:

 \begin{equation}\label{Trigger1}
  \left(D_1 + D_2\right) \times \left(D_3 + D_4\right)
\end{equation}
 in the $p + \mathrm{C}$ and $d + \mathrm{C}$ experiments
 and with the additional requirement of
anticoincidence with the signals from the scintillation counters in the $d + \mathrm{Cu}$ experiment:

\begin{equation}\label{Trigger2}
 \left(D_1 + D_2\right) \times \left(D_3 + D_4\right) \times \overline{S1}
\times \overline{S2}.
\end{equation}
In realizing the trigger conditions the amplitudes of all
32 modules were recorded on a disc. The dead time of data acquisition is about
150 $\mu$s per trigger.
%The mean rate of triggering was about $2.6 \times 10^3$ events per spill.
The duration of the irradiation cycle is 1 s.
%Totally about $0.7 \times 10^6$ triggers were recorded during this experiment.

The data presented were collected in experiments to study
production of the $\eta$-meson, so a coincidence of both arms of the
spectrometer was required for triggering.  At the request of \cite{van Beveren
Rupp private communication} we analyzed the recorded data for an excess above
background of coincidences in a single arm of the spectrometer.  The requirement
of coincidence of both arms reduced the detection efficiency for this purpose,
(to about $2 \times 10^{-7}$), but due to the high collected statistics,
(about $2\times 10^{12}$ $d + \mathrm{C}$ interactions, $10^{11}$ $p + \mathrm{C}$ interactions, and $0.8\times 10^{12}$ $d + \mathrm{Cu}$ interactions),
it was possible to observe a significant excess.

\subsection{Event selection}
\label{sec: Event selection}

Photons in the detector are recognized as isolated and confined clusters (an
area of adjacent modules with a signal above the threshold) in the
electromagnetic calorimeter. The photon energy is calculated from the energy of
the cluster. Cluster characteristics were tested by comparison with Monte-Carlo
simulations of electron-photon showers in Cherenkov counters by means of the
program package EMCASR \cite{EMCASR}. The results obtained earlier with
extracted ion beams of the JINR Synchrophasotron have demonstrated a good
agreement between experimental and simulated data \cite{Ab94}. Assuming that the
photon originates from the target, its direction is determined from the
geometrical positions of constituent modules weighted with the corresponding
energy deposit in activated modules.

After an analysis of the individual modules (see Appendix 1) and exclusion of some modules because of their poor performance, (6 modules in the left arm and one module in the right arm of the spectrometer were excluded), the data were processed by an event reconstruction program and were recorded on DST.
As a result, about $0.3 \times 10^6$ events were recorded under the following condition:
the number $N_\gamma$ of detected photons in an event with energy $E_\gamma > 50$ MeV
is $N_\gamma \ge 2 $, such that there are 1 or more photons in each arm \cite{R1}.

To search for a signal at the low effective masses we have analyzed photon pairs
detected in the same arm of the $\gamma$-spectrometer. Below are results of this
analysis for photon pairs detected in the Right arm of the $\gamma$-spectrometer
(situated at an angle of 26$^\circ$, see Fig.\hspace{-0.3ex} \ref{Experimental
instrumentation}).

\begin{figure}[htb]
   \centering
    \subfigure{
      \includegraphics[width=0.489\textwidth]{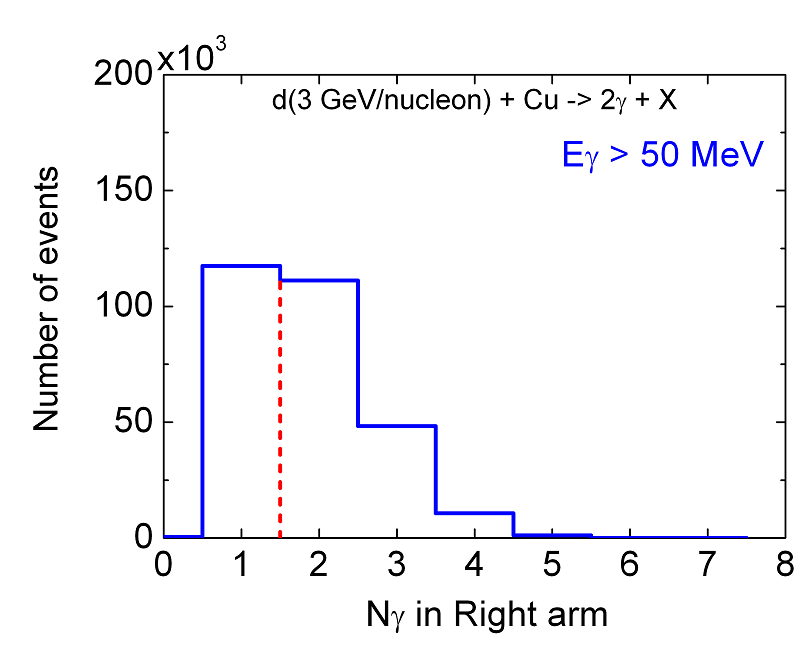}}
\hspace{-0.45cm}
    \subfigure{
      \includegraphics[width=0.489\textwidth]{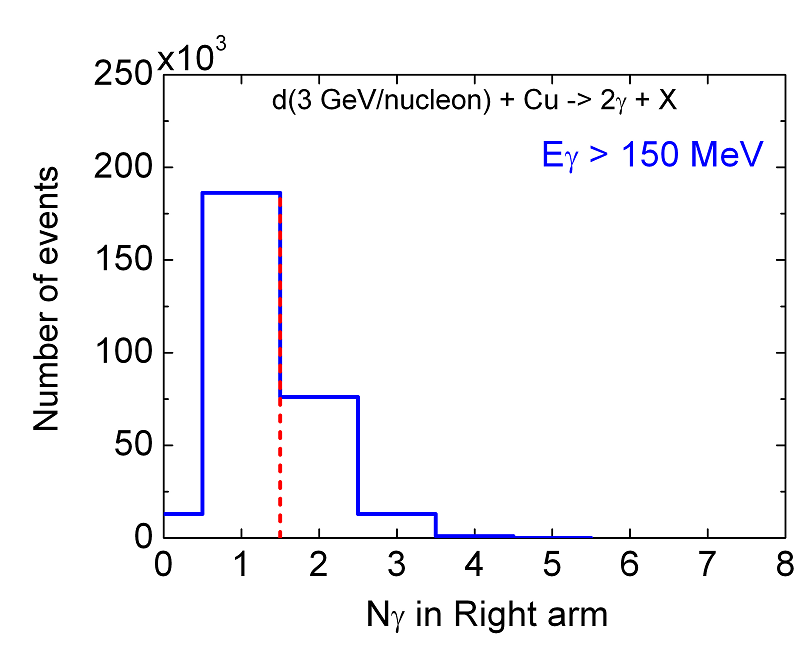}}
\caption{Numbers of detected photons in the Right arm of the
$\gamma$-spectrometer at a minimal energy of photons of 50 MeV (left figure) and
150 MeV (right figure).}
\label{Multiplicities of detected photons}
\end{figure}

\textbf{The multiplicities} of detected photons in the said arm at
different energy selection levels in the $d + \mathrm{Cu}$ experiment are shown in Fig.\hspace{-0.6ex}
\ref{Multiplicities of detected photons}.

% MultiplicitiesOfDetectedPhotonsA.png MultiplicitiesOfDetectedPhotonsB.png

Totally, there are $1.7 \times 10^5$ and $0.9 \times 10^5$ events with two or
more photons detected in the Right arm, with a minimum energy of photons of 50
MeV and 150 MeV respectively.
\vspace{0.4ex}

% Insertion 1
\textbf{The expected width} of the signal is defined by the uncertainties in the measured photon energies and in the opening angle of the photons. The errors of measured opening angles were estimated from the analysis of simulated data, and are as follows: standard deviation, $\Delta\Theta\leq 0.9 ^\circ$ ($\Delta\Theta \simeq 0.9 ^\circ$ in cases where each photon triggers just one unit of the spectrometer, because the coordinates of a photon hit are determined more accurately, when the photon triggers several modules).
The uncertainties $\Delta E_\gamma$ of measured values of photon energies were estimated by the empirical formula for energy resolution averaged over the surface of the spectrometer \cite{Kha85, Ab89}:
\begin{equation}\label{Resolution}
  \Delta E_\gamma /E_\gamma \approx 0.068/\sqrt{E_\gamma },
\end{equation}
where $E_\gamma$ is in GeV.

Expected width $\Delta M_{\gamma\gamma}$ of the effective mass of photon pairs were evaluated by the formula (for the considered region of the opening angles ($\Theta\leq 16 ^\circ$) $M \approx\sqrt{E_1 E_2}\cdot\Theta$):
\begin{equation}\label{Mass_resolution}
(\Delta M_{\gamma\gamma} /M_{\gamma\gamma} )^2 =(\Delta\Theta /\Theta)^2 + (\frac{1}{2}\Delta E_1 /E_1)^2+(\frac{1}{2}\Delta E_2 /E_2)^2 .
\end{equation}

\begin{figure}[htb]
   \centering
    \subfigure{
      \includegraphics[width=0.49\textwidth]{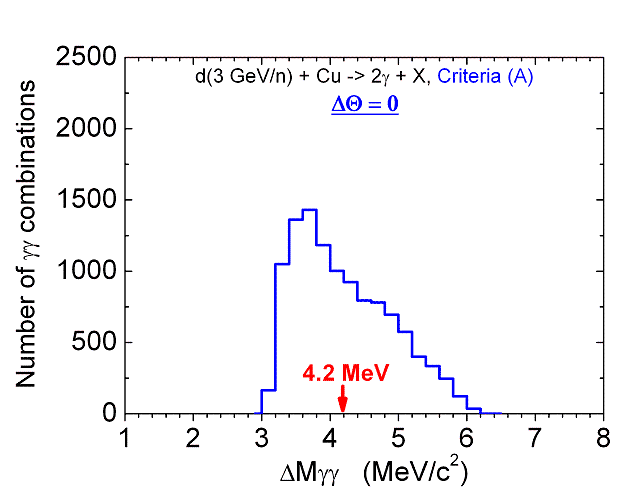}}
\hspace{-0.45cm}
    \subfigure{
      \includegraphics[width=0.49\textwidth]{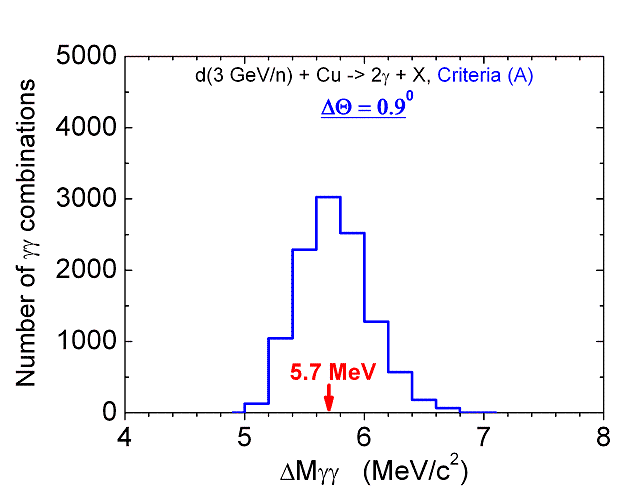}}
\caption{Distributions of calculated value of $\Delta M_{\gamma\gamma}$ for photon pairs selected from the Right arm of the $\gamma$-spectrometer, under the Criteria (A), for the vicinity of the E(38) mass: $36 < M_{\gamma\gamma} < 40$ MeV/$c^2$, at $\Delta\Theta = 0$ (left figure) and at the maximum value of $\Delta\Theta = 0.9 ^\circ$ (right figure).}
\label{DeltaMdT0 and DeltaMdT09}
\end{figure}

To estimate $\Delta M_{\gamma\gamma}$ we have built a distribution of calculated value of this quantity for the photon pairs detected in the experiment, with an effective mass in the ranges of 16-18 and 36-40 $MeV/c^2$ (in the vicinities of the $X17$ and $E(38)$ masses).
The
results at $\Delta\Theta = 0$ and at the maximum value of $\Delta\Theta = 0.9 ^\circ$ for the range 36-40 $MeV/c^2$
are shown in Fig.\hspace{-0.3ex} \ref{DeltaMdT0 and DeltaMdT09}.
Taking into account the various configurations of clusters (containing 2 or 3 adjacent modules, when the uncertainties of the measured values of the opening angles are less than $0.9^\circ$) we have obtained
  the average values of the width of the signals under the criteria (i)-(v) (see below, Fig.\hspace{-0.3ex} \ref{Criteria A spectrum}) and
 under the Criteria (A) ($E_{\gamma} > 50$ MeV, $450 < E_{12} < 750$ MeV \cite{Ab2019}):
$\langle \Delta M_{\gamma\gamma}\rangle \approx
3 \textrm{ MeV}/c^2$ and $\langle \Delta M_{\gamma\gamma}\rangle \approx
5 \textrm{ MeV}/c^2$ respectively.

% DeltaMdT0 and DeltaMdT09.png

In order to identify the signal from detected particles all photon pair
combinations are used to calculate the invariant mass in each event.

To see a possible structure of the invariant mass spectra, a background should
be subtracted. The so-called event mixing method was used to estimate the
combinatorial background: a photon in one event from a group of modules
is combined
with photon in other events from the same group. In the mixing there are involved events
in which there are two or more photons in the group satisfying the selection criteria.
This background was subtracted from the invariant mass distributions (see
 bottom panels in Fig.\hspace{-0.2ex} \ref{Criteria A spectrum}).

\subsection{Optimal conditions for X17}
\label{sec: Optimal conditions for X17}

In order to study the region of small invariant masses, we processed the data obtained
in groups not participating in the trigger launch (thanks to the logical addition (see (\ref {Trigger1}), (\ref {Trigger2})), there are such groups in each event).
To collect sufficient statistics, we processed the data obtained in several experiments.

\begin{figure}[p]
   \centering
    \subfigure{
      \includegraphics[width=0.493\textwidth]{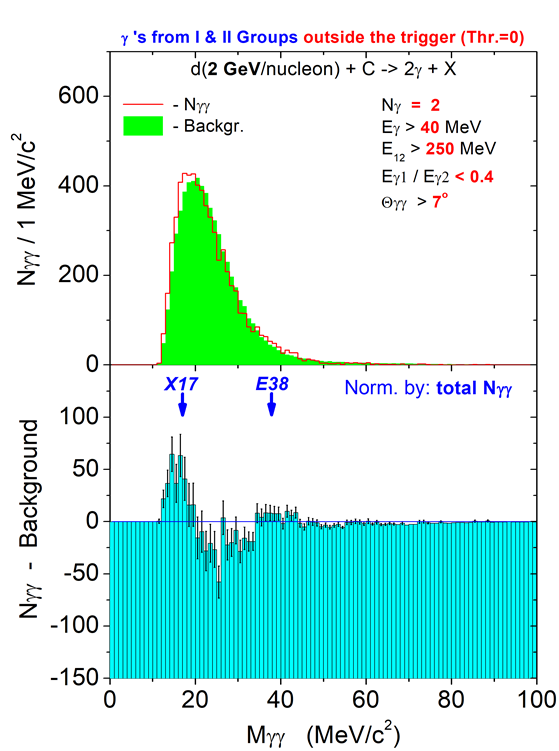}}
\hspace{-0.4cm}
    \subfigure{
      \includegraphics[width=0.493\textwidth]{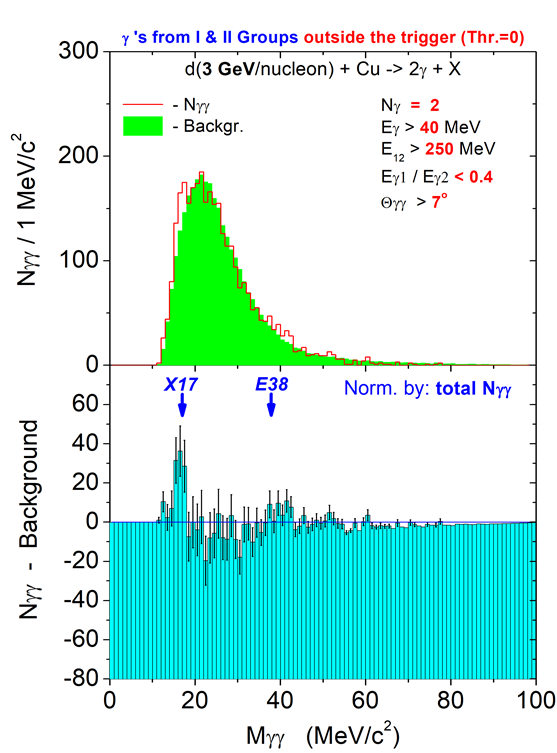}}
    \subfigure{
      \includegraphics[width=0.493\textwidth]{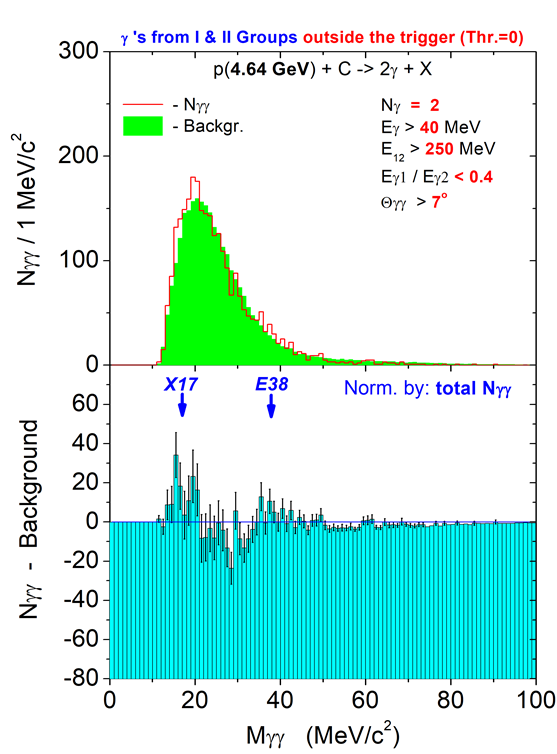}}
\hspace{-0.4cm}
    \subfigure{
      \includegraphics[width=0.493\textwidth]{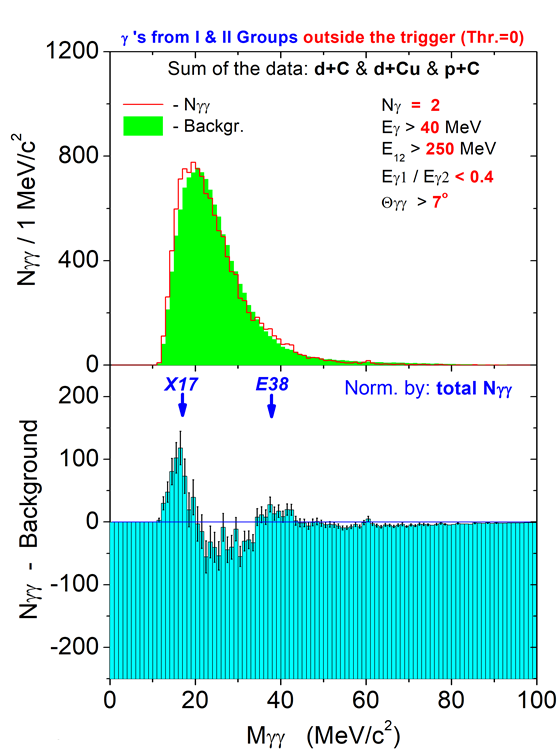}}
\caption{Invariant mass distributions of $\gamma \gamma$ pairs satisfying
criteria (i)-(v) without (upper panels) and with (bottom panels) the
background subtraction obtained for the $d +
\mathrm{{{C}}}$, $d + \mathrm{{{Cu}}}$ and $p + \mathrm{{{C}}}$ reactions. The backgrounds are normalized to the total pair numbers
in the spectra.}
\label{Criteria A spectrum}
\end{figure}

 Figure \ref{Criteria A spectrum} shows the invariant mass distributions
 of $\gamma \gamma$ pairs under the optimal conditions for searching for a
 particle with a mass of 17 MeV/$c^2$:
\begin{quote}
(i) the number of detected photons in the group, $N_\gamma = 2$;\\
(ii) the minimal energy of photons, $E_{\gamma Min} = 40$ MeV;\\
(iii) the sum of the energies of photons in a pair, $E_{12} > 250$ MeV
(effective detection of pairs at the setup geometry);\\
(iv) the ratio of the energies $E_{\gamma 1}/E_{\gamma 2} < 0.4$
(suppresses systematic errors due to violation of the energy-momentum conservation laws at the event mixing);\\
(v) the opening angles of photons in a pair, $\Theta_{\gamma \gamma} >
7^{o} $.
\end{quote}

% Part4_1(a)-(d).png

Figure \ref{Criteria A spectrum} shows the sum of data for two groups that did not participate in the event triggering (after logical addition). Thus, the energy in the specified group could be arbitrary (without the influence of the discriminator thresholds).

\begin{figure}[p]
   \centering
    \subfigure{
      \includegraphics[width=0.493\textwidth]{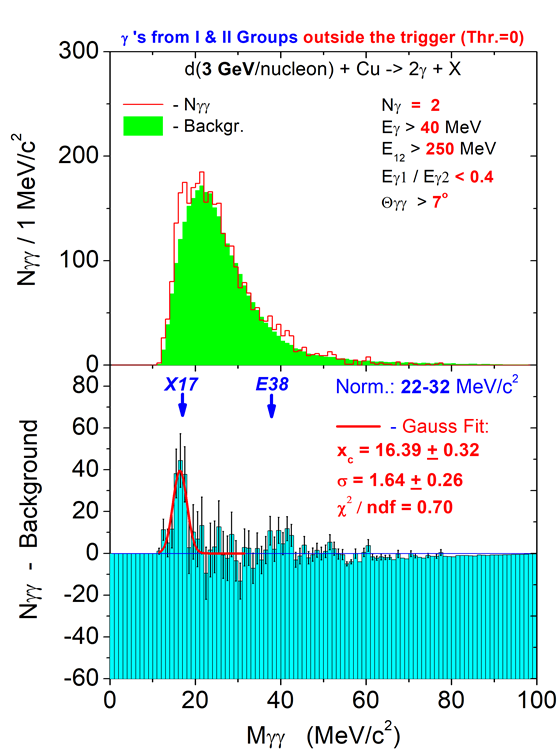}}
\hspace{-0.4cm}
    \subfigure{
      \includegraphics[width=0.493\textwidth]{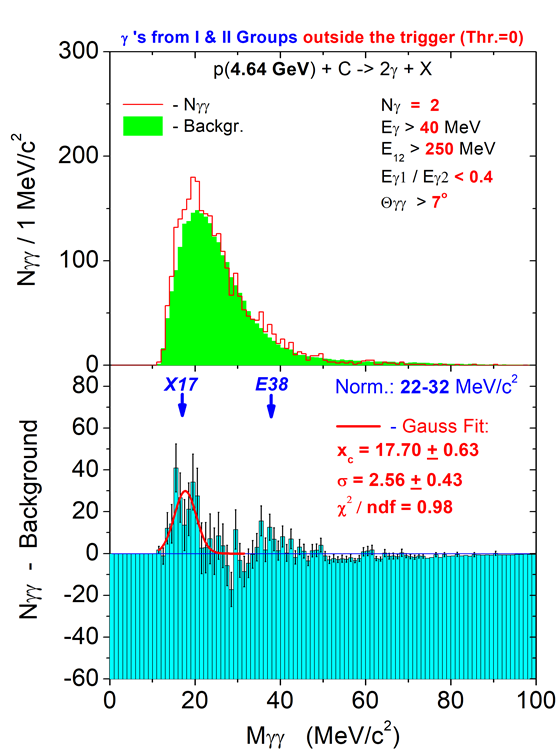}}
    \subfigure{
      \includegraphics[width=0.493\textwidth]{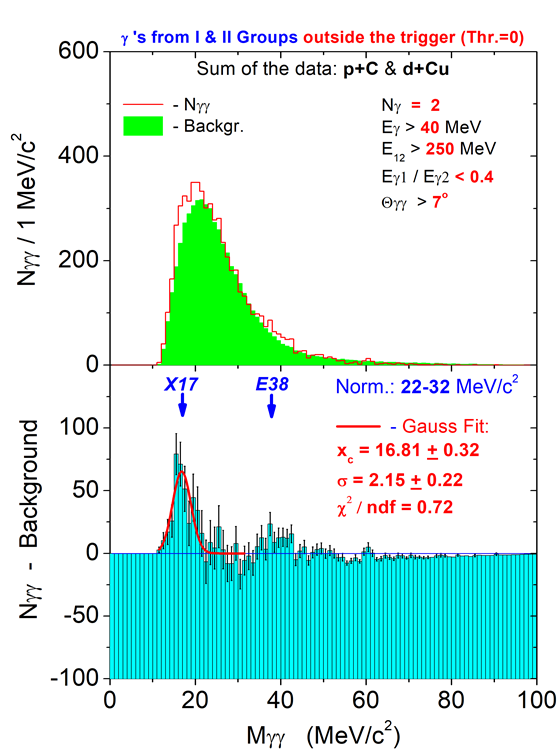}}
\hspace{-0.4cm}
    \subfigure{
      \includegraphics[width=0.493\textwidth]{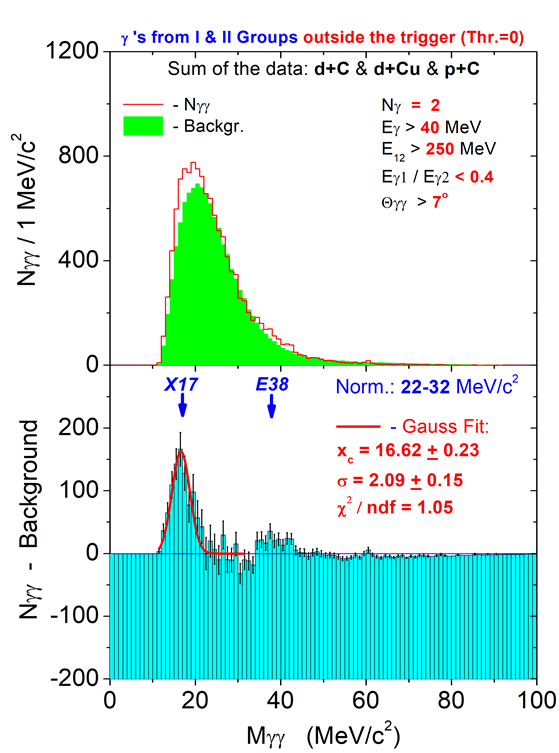}}
\caption{The same as in Fig.\hspace{-0.2ex} \ref{Criteria A spectrum}, but after the backgrounds normalization by
the numbers of pairs in the range (22, 32) MeV$/c^2$.}
\label{Criteria A Norm}
\end{figure}

The same as in Fig.\hspace{-0.2ex} \ref{Criteria A spectrum}, but after the normalization of the background
by the number of pairs in the nonresonant region of 22–32 MeV, are shown in Fig.\hspace{-0.2ex} \ref{Criteria A Norm}.

The curve in Fig.\hspace{-0.2ex} \ref{Criteria A Norm} is the Gaussian
approximation of the experimental points in the range (11; 32) MeV/$c^2$:

\begin{equation}
\label{Gaussian approximation}
 \frac{dN}{dM_{\gamma \gamma}} = \frac{N_0}{\sigma \cdot \sqrt{2\pi}} \cdot
\mathrm{exp}\left(-\frac{\left(M_{\gamma \gamma} - \textsl{x}_c \right)^2}{2\sigma^2}
\right).
\end{equation}

The number of $\gamma \gamma$ pairs in the range of 12-22 MeV/$c^2$
after the background subtraction in the sum of three experiments
is $924 \pm 77$.
The values of the obtained fitting parameters in (\ref{Gaussian approximation})
are in the pictures. The parameter $N_0$ for the sum of the data obtained in the $p + \mathrm{C}$, $d + \mathrm{C}$ and $d + \mathrm{Cu}$ experiments, is:
\vspace{-1.2ex}
\begin{center}
$N_0=856\pm 75$.
\end{center}
\vspace{-1.2ex}
Thus, the statistics in the observed structure
about 17 MeV/$c^2$ is more than 11 standard deviations.
Based on the changes of the signal position ($\textsl{x}_c$ parameter) in the different experiments (from 16.4 to 17.7 MeV/$c^2$),
we estimate the possible systematic errors to be no more than $\pm 0.7$ MeV/$c^2$.

\subsection{Gamma's from the triggering Groups}
\label{sec: Gamma's from the triggering Groups}

\begin{figure}[p]
   \centering
    \subfigure{
      \includegraphics[width=0.493\textwidth]{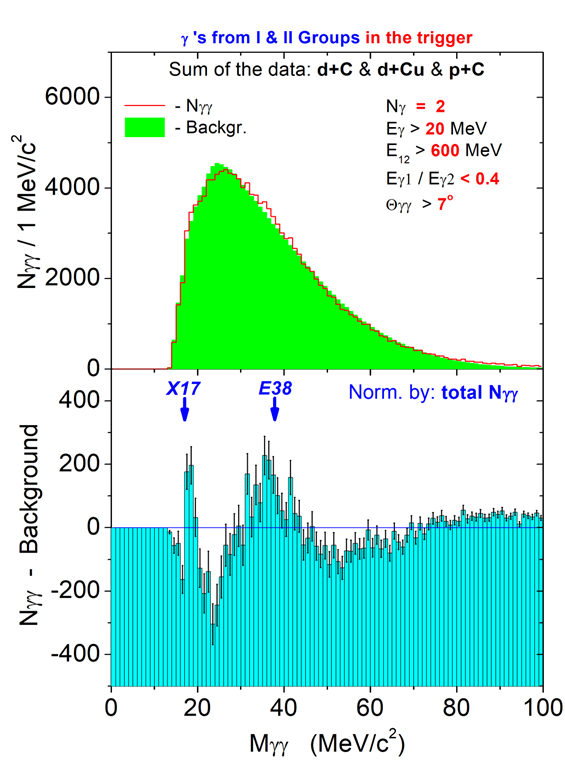}}
\hspace{-0.4cm}
    \subfigure{
      \includegraphics[width=0.493\textwidth]{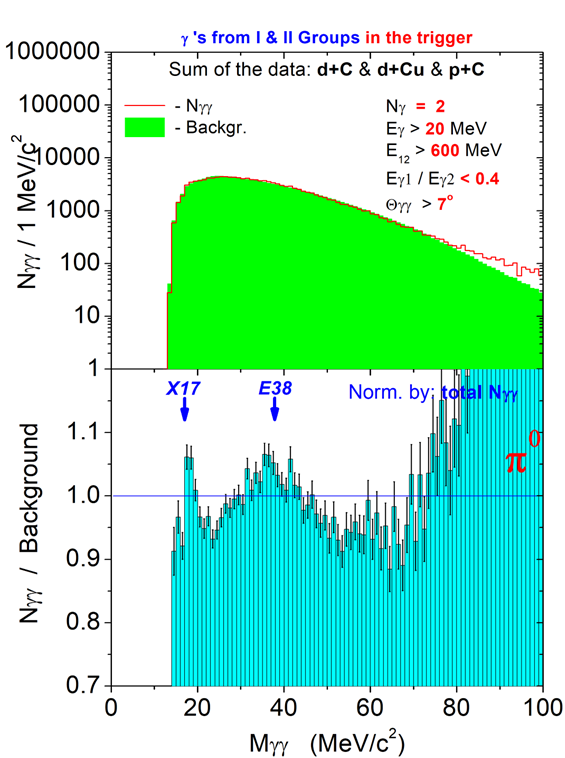}}
\caption{Invariant mass distributions of $\gamma \gamma$ pairs satisfying
criteria (1)-(5) without (upper panels) and with (bottom panels) the
background subtraction obtained for the $d +
\mathrm{{{C}}}$, $d + \mathrm{{{Cu}}}$ and $p + \mathrm{{{C}}}$ reactions. The backgrounds are normalized to the total pair numbers
in the spectra.}
\label{IGrL1etc}
\end{figure}

In our recent works \cite{Ab2019, Abraamyan et al Aug 12} we reported the results of an analysis of the spectra of photon pairs detected in the same arm of the spectrometer (the hodoscope including the above two groups). As a result of this analysis, a statistically reliable signal was found at an invariant mass
of about 38 MeV/$c^2$.
Below are the results of an analysis of photon pairs detected only in one group participating in the launch of the facility. Thus, the sum of photon energies in these events are influenced by the discriminator thresholds (see expressions (\ref{Trigger1}) and (\ref{Trigger2})).
Figure \ref{IGrL1etc} shows
the invariant mass distributions
 of $\gamma \gamma$ pairs under the following conditions:
\begin{quote}
(1) the number of detected photons in the group, $N_\gamma = 2$;\\
(2) the minimal energy of photons, $E_{\gamma Min} = 20$ MeV;\\
(3) the sum of the energies of photons in a pair, $E_{12} > 600$ MeV
(taking into account the discriminator thresholds);\\
(4) the ratio of the energies $E_{\gamma 1}/E_{\gamma 2} < 0.4$
(suppresses systematic errors due to violation of the energy-momentum conservation laws at the event mixing);\\
(5) the opening angles of photons in a pair, $\Theta_{\gamma \gamma} >
7^{o} $.
\end{quote}

\section{CHECK OF THE OBSERVED PEAKS}
\label{sec: CHECK THE OBSERVED PEAK}

Systematic errors may be due to uncertainty in measurements of $\gamma$ energies
and inaccuracy in estimates of the combinatorial background. The method of
energy reconstruction of events is described in detail in Refs. \cite{Ab89, Ab94}.
Possible overlapping effects were investigated previously for the
reaction with the higher masses of the colliding nuclei and at higher energies –
in the reaction of C + C at 4.5 GeV/$c$ per nucleon \cite{Ab94}. It was found
that the average displacement of the effective masses of $\gamma \gamma$-pairs
in the reaction is only 6\%. Thus, the influence of the overlap in the present
experiment is negligible.

\subsection{Different minimal energies of photons}
\label{sec: Diff Eg}

\begin{figure}[p]
   \centering
    \subfigure{
      \includegraphics[width=0.493\textwidth]{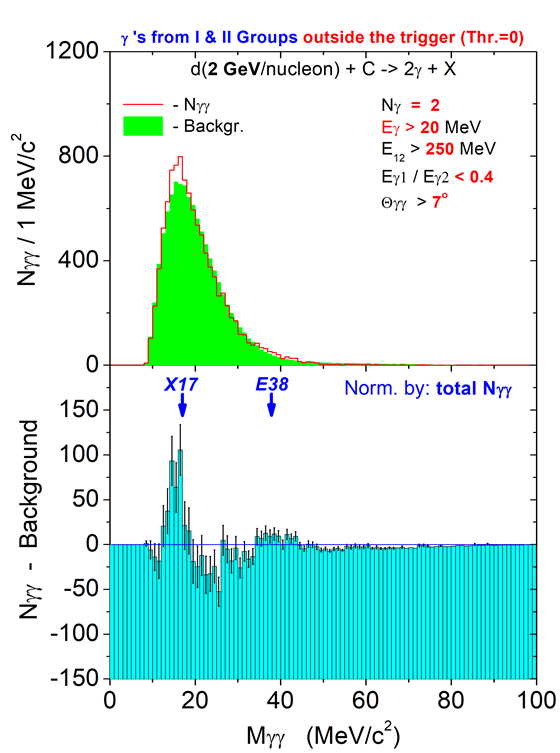}}
\hspace{-0.4cm}
    \subfigure{
      \includegraphics[width=0.493\textwidth]{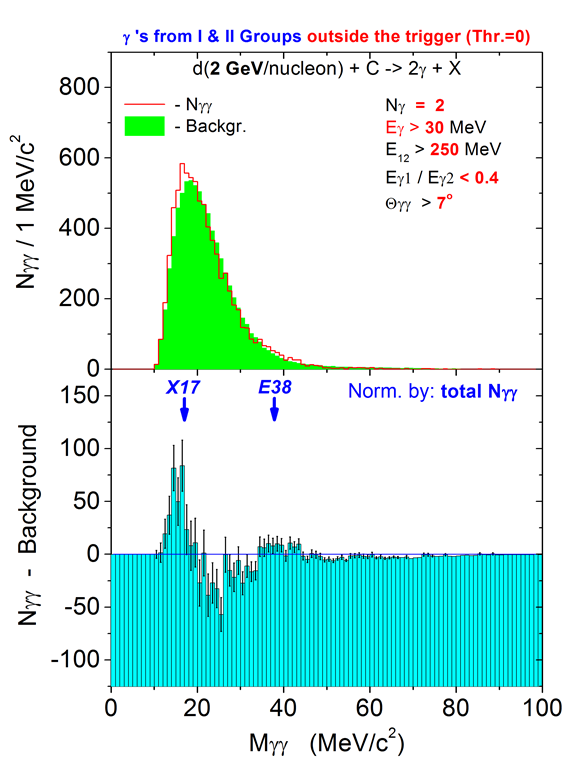}}
    \subfigure{
      \includegraphics[width=0.493\textwidth]{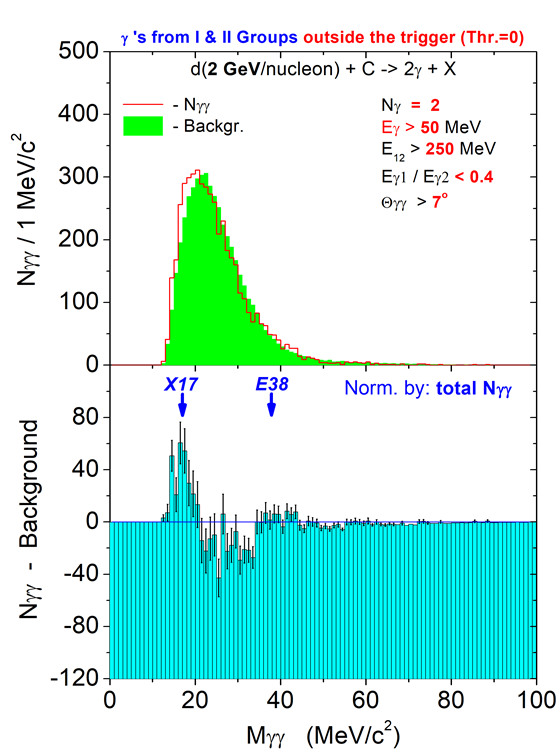}}
\hspace{-0.4cm}
    \subfigure{
      \includegraphics[width=0.493\textwidth]{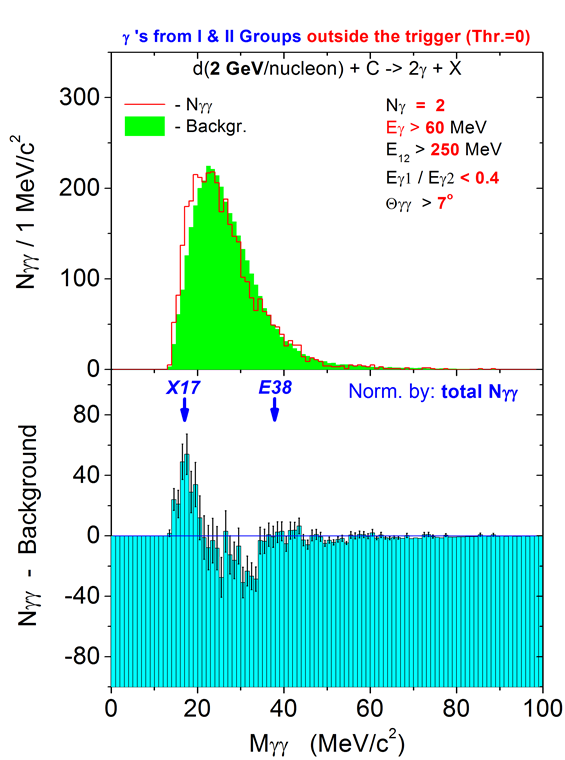}}
\caption{The same as in Fig.\hspace{-0.2ex} \ref{Criteria A spectrum}, but at
different minimal energies of photons.}
\label{Criteria A Diff Eg}
\end{figure}

One way to check the signal is to change the minimal photon energy, which significantly changes the background position.
Fig.\hspace{-0.2ex} \ref{Criteria A Diff Eg} shows the same spectra as in Fig.\hspace{-0.2ex} \ref{Criteria A spectrum}, but at different minimal photon energies.

\subsection{Gamma's from different Groups. Signals from the \texorpdfstring{\(\pi^0\)}{pi0}-mesons}
\label{sec: Diff Gr}

One of the criteria of accuracy of energy reconstruction is the conformity of
the peak positions corresponding to the known particle mass values. As was shown
in \cite{R1} the position of peaks corresponding to $\eta$- and $\pi^0$-mesons
in these experiments (at selection the photons from different arms of the
spectrometer) is in reasonable agreement with the table values of their masses.

\begin{figure}[p]
   \centering
    \subfigure{
      \includegraphics[width=0.493\textwidth]{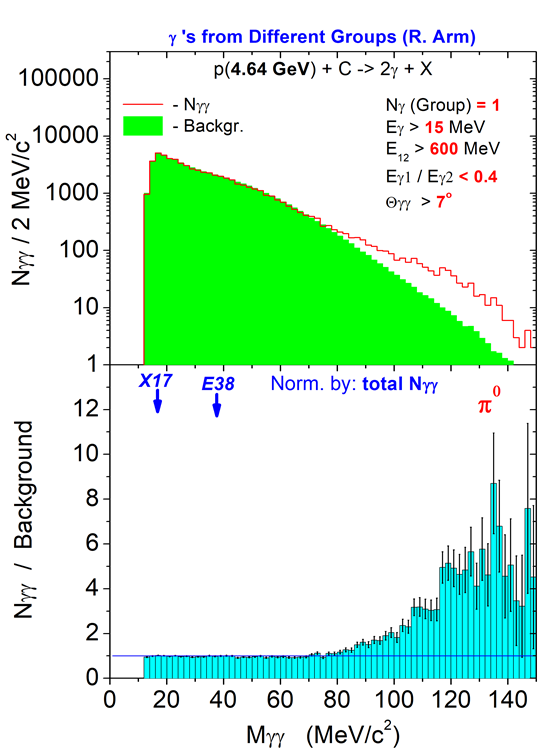}}
\hspace{-0.4cm}
    \subfigure{
      \includegraphics[width=0.493\textwidth]{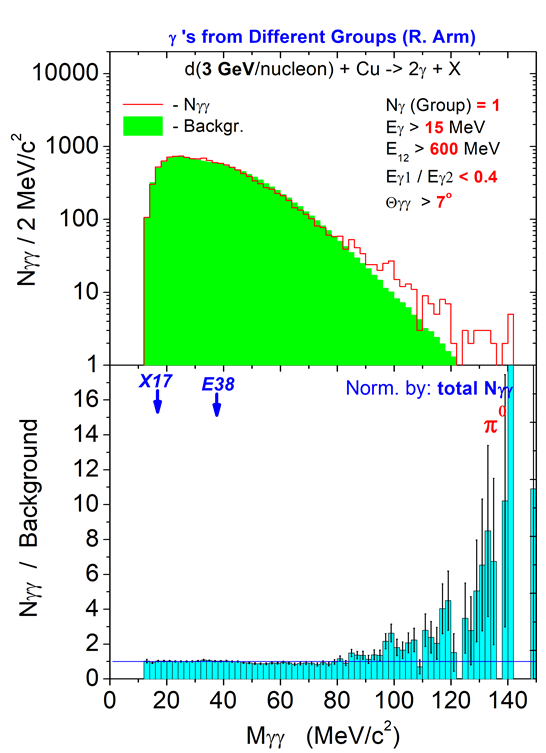}}
    \subfigure{
      \includegraphics[width=0.493\textwidth]{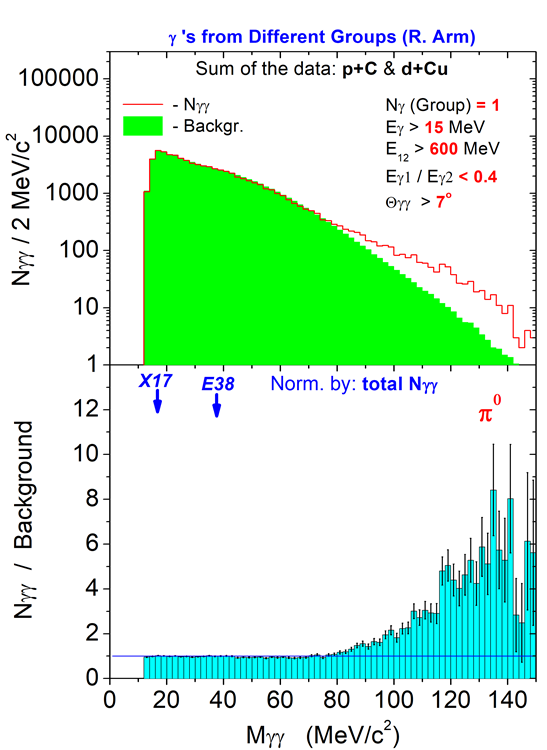}}
\hspace{-0.4cm}
    \subfigure{
      \includegraphics[width=0.493\textwidth]{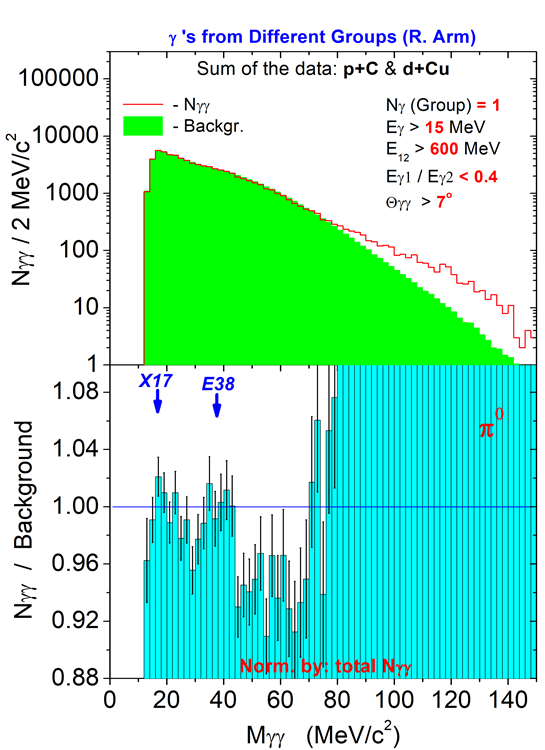}}
\caption{The $N_{\gamma \gamma}$ to the background ratio for photon pairs from different groups (Right arm) for
the $d + \mathrm{{{Cu}}}$ and $p + \mathrm{{{C}}}$ reactions (upper figures) and sum of these data at different scales.}
\label{Criteria B Diff Gr}
\end{figure}

For the simultaneous observation of signals from \texorpdfstring{\(\pi^0\)}{pi0}-mesons, it is necessary to select pairs from different groups to provide a wider range of opening angles.
With such a selection, systematic errors in the background (obtained by the event mixing method) associated with thresholds on the sum of energies in groups can arise
(at least in one of the two groups, the sum of energies is higher than the threshold). The thresholds can be taken into account via cuts of the sum of the
energies of photons in a pair. The $N_{\gamma \gamma}$ to the background ratio for photon pairs from different groups (Right arm) is shown in Fig.\hspace{-0.2ex} \ref{Criteria B Diff Gr}.

\section{DATA SIMULATION}
\label{sec: DATA SIMULATION}

To simulate the $d + \mathrm{Cu}$ reaction we use a transport code. At high
energies it is the Quark-Gluon String Model (QGSM) \cite{QGSM} and at the energy
of a few GeV the string dynamics is reduced to the earlier developed Dubna
Cascade Model (DCM) \cite{Dub_casc} with upgrade of elementary cross sections
involved \cite{LAQGSM}.

The DCM divides the collision into three stages, well separated in time. During
the initial stage an intranuclear cascade develops, primary particles can
scatter and secondary particles can re-scatter several times prior to their
absorption or escape from the nucleus. At the end of this step the coalescence
model is used to localize $d$, $t$, $^3$He, and $^4$He particles from nucleons
found inside spheres with well-defined radii in configuration space and momentum
space. The emission of cascade particles determines a particle-hole
configuration, i.e., Z, A, and excitation energy that is taken as the starting
point for the second, pre-equilibrium stage of the reaction, described according
to the cascade exciton model \cite{exciton}. Some pre-equilibrium particles may
be emitted and this leads to a lower excitation of the thermalized residual
nuclei. In the third, final evaporation/fission stage of the reaction, the
de-excitation of the residue is described with the evaporation model. The last
two stages are important for triggering the events. All components contribute
normally to the final spectra of particles and light fragments; low-energy
evaporated photons are not included into subsequent analysis. For relativistic
energies the cascade part of the DCM is replaced by the refined cascade model,
which is a version of the quark-gluon string model (QGSM) developed in
\cite{Toneev90} and extended to intermediate energies in \cite{Amelin90}. The
description of the mean-field evolution is simplified in the DCM in the sense
that the shape of the scalar nuclear potential, defined by the local
Thomas-Fermi approximation, remains the same throughout the collision. Only the
potential depth changes in time, according to the number of knocked-out
nucleons. This frozen mean-field approximation allows us to take into account
the nuclear binding energies and the Pauli exclusion principle, as well as to
estimate the excitation energy of the residual nucleus by counting the excited
particle-hole states. This approximation is usually considered to work
particularly well for hadron-nucleus collisions.

The following $\gamma$-decay channels are taken into account: the direct decays
of $\pi^0$, $\eta$, $\eta'$ hadrons into two $\gamma$'s, $\omega \to \pi^0
\gamma$, $\Delta \to N \gamma$ and the Dalitz decay of $\eta \to \pi^+ \pi^-
\gamma$, $\eta \to \gamma + e^+ + e^-$ and  $\pi^0 \to \gamma + e^+ + e^-$, the
$\eta' \to \rho^0 + \gamma$, the $\Sigma \to \Lambda + \gamma$, the $\pi N$ and
$N N$-bremsstrahlung. One should note that in accordance with the HADES data
\cite{HADES}, the $p n$-bremsstrahlung turned out to be higher by a factor of
about 5 than a standard estimate and weakly depends on the energy. This finding,
being in agreement with the result of Ref.\hspace{-0.5ex} \cite{Copt}, allowed
one to resolve the old DLS puzzle \cite{Brat07}. This enhancement factor is
included in our calculations. Tests of this model in detail are described in
Ref.\hspace{-0.5ex} \cite{R1}.

\subsection{Estimates of systematic errors in the combinatorial background}
\label{sec: Systematic errors in the background}

\begin{figure}[p]
   \centering
    \subfigure{
      \includegraphics[width=0.453\textwidth]{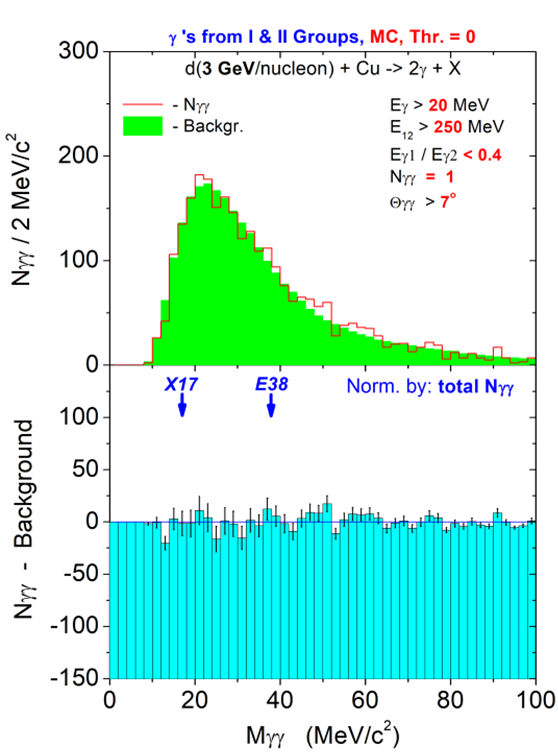}}
\hspace{-0.43cm}
    \subfigure{
      \includegraphics[width=0.453\textwidth]{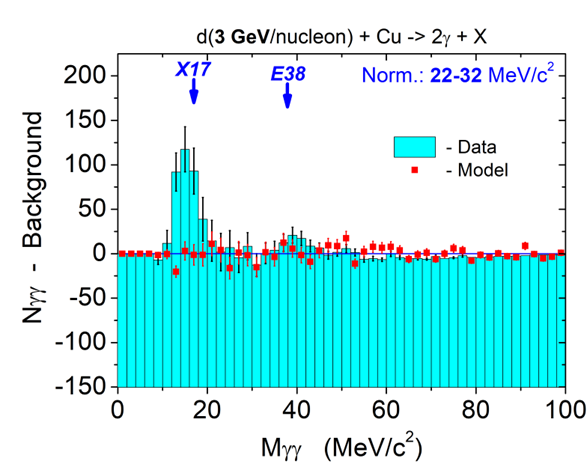}}
\caption{The result of processing the simulated data by the same method as experimental data, for the $d$(3 GeV/n) $+\mathrm{Cu}$ reaction
(left) and comparison of the experimental and simulated spectra after the backgrounds subtraction (right).
The backgrounds are normalized
by the numbers of pairs in the range (22, 32) MeV$/c^2$.
The selection criteria are in the figure.}
\label{Model without E38}
\end{figure}

The results of processing the simulated data by the method which was applied to
the experimental data (i.e., the difference of the combinatorial spectrum and
the background obtained by
the event mixing method, see Fig.\hspace{-0.2ex} \ref{Criteria A spectrum}) are shown in
Fig.\hspace{-0.2ex} \ref{Model without E38}.

% ModelWithoutE38FigA.png ModelWithoutE38FigB.png

As seen from Fig.\hspace{-0.4ex} \ref{Model without E38}, there is no
structure in the vicinity of the $X17$ and $E(38)$ masses.

\begin{figure}[p]
   \centering
    \subfigure{
      \includegraphics[width=0.493\textwidth]{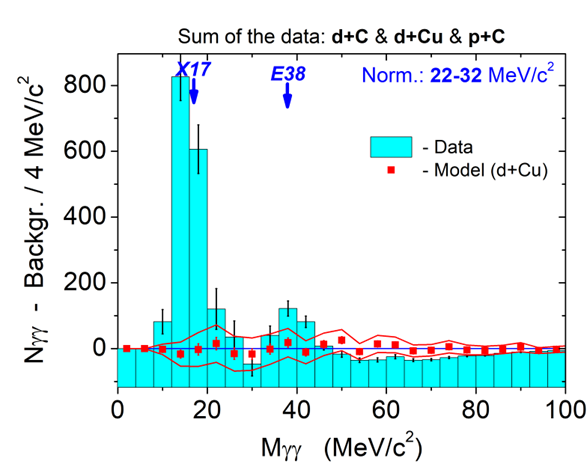}}
\hspace{-0.43cm}
    \subfigure{
      \includegraphics[width=0.497\textwidth]{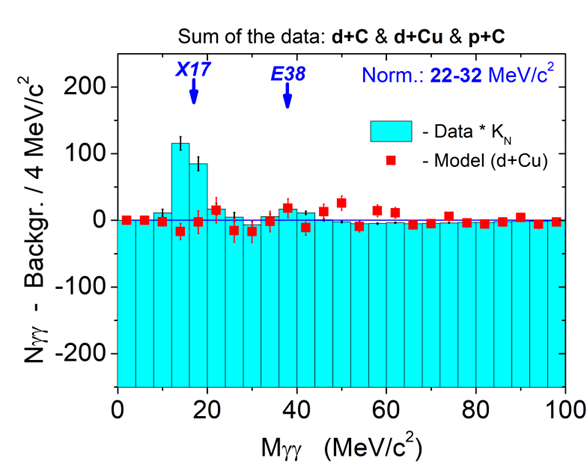}}
\caption{The same as in the Fig.\hspace{-0.4ex} \ref{Model without E38} (right figure), but for the sum of the spectra (after the backgrounds subtraction), obtained in three experiments (indicated in the figure). The curves indicate the interval of $\pm$ 3 standard statistical errors in the simulated data. On the right figure - the same, but after reducing the experimental data (see the text, Eq.\hspace{-0.2ex} \ref{KN}).}
\label{QuantCheck}
\end{figure}

For a quantitative check of the signals,
the result of processing the simulated data were compared with the sum of the spectra
(after the background subtraction), obtained in three experiments (see
Fig.\hspace{-0.2ex} \ref{QuantCheck}).
The right figure of Fig.\hspace{-0.2ex} \ref{QuantCheck} shows the most stringent verification of the signals: the experimental data reduced to the simulated data, i.e. the accounts in the experimental data (in the left figure) were multiplied by a factor equal to the ratio of the numbers of $\gamma \gamma$ pairs
in the range (22, 32) MeV$/c^2$:
\begin{equation}
\label{KN}
K_N=N^{Model}_{\gamma \gamma}(22<M_{\gamma \gamma}<32\; MeV/c^2)\, /\, N^{Exper.}_{\gamma \gamma}(22<M_{\gamma \gamma}<32\; MeV/c^2).
\end{equation}
As seen  from the figure, the signal at an invariant mass of $\sim$17 MeV$/c^2$ is statistically significant.
A more rigorous quantitative verification of the signal at $\sim$38 MeV$/c^2$ was given in
Ref.\hspace{-0.5ex} \cite{Ab2019}.

% ModelWithoutE38FigA.png ModelWithoutE38FigB.png

\begin{figure}[p]
   \centering
    \subfigure{
      \includegraphics[width=0.493\textwidth]{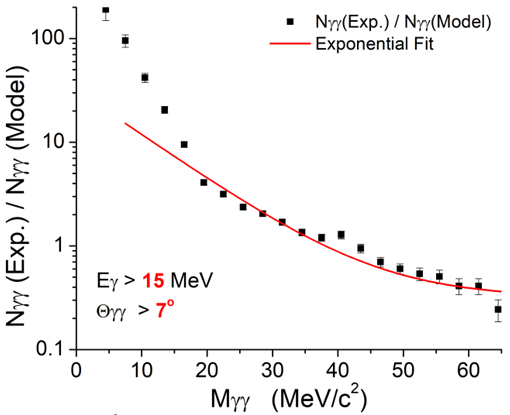}}
\hspace{-0.43cm}
    \subfigure{
      \includegraphics[width=0.497\textwidth]{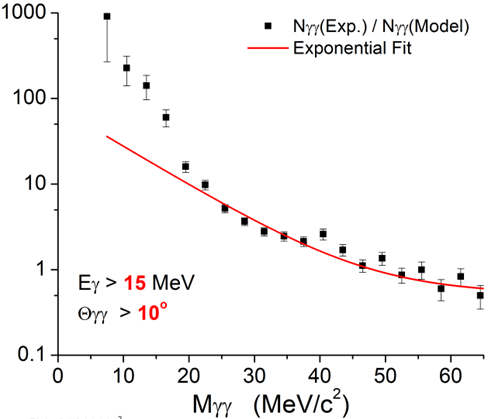}}
\caption{The ratio of experimental and simulated invariant mass spectra at the minimum levels of
energy cuts and at two levels of opening angle cuts, for the reaction $d + \mathrm{Cu}\rightarrow \gamma+\gamma+x$ at 3.8 GeV/c per nucleon.}
\label{EnaM}
\end{figure}

The ratio of experimental and simulated data is shown in Fig.\hspace{-0.2ex} \ref{EnaM}. As seen from the figure, in addition to the significant difference between the indicated spectra in the low-mass region (which is a consequence of both the high thresholds applied in the experiment and the known excess of soft photons), a statistically significant excess in the invariant mass region of about 38 MeV$/c^2$ is also seen.

% ModelWithoutE38FigA.png ModelWithoutE38FigB.png

\subsection{No cuts. Systematic errors}
\label{sec: No cuts}

\begin{figure}[p]
   \centering
    \subfigure{
      \includegraphics[width=0.493\textwidth]{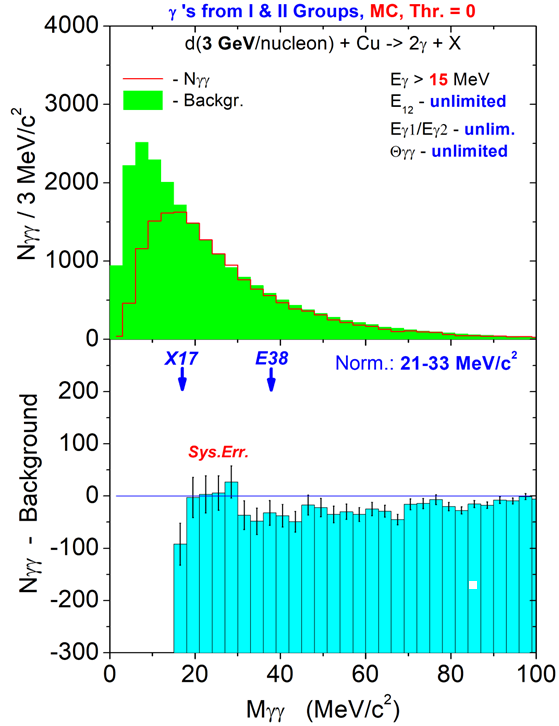}}
\hspace{-0.43cm}
    \subfigure{
      \includegraphics[width=0.497\textwidth]{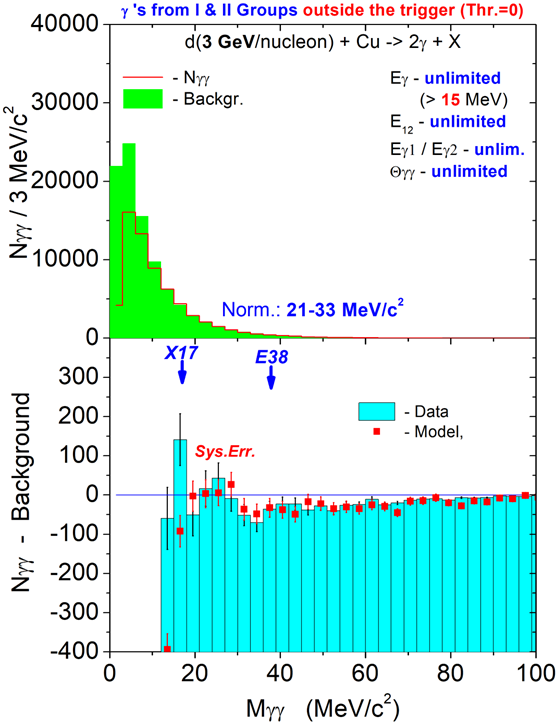}}
\caption{The same as in Fig.\hspace{-0.2ex} \ref{Model without E38}, but at the minimum selection conditions,
for the simulated (left figure) and experimental data (right figure)
for the reaction
$d + \mathrm{Cu}\rightarrow \gamma+\gamma+x$ at 3.8 GeV/c per nucleon. The backgrounds are normalized
by the numbers of pairs in the range (21, 33) MeV$/c^2$}
\label{No cuts T0}
\end{figure}

To determine the effect of the applied selection criteria, the spectra presented in Figs.\hspace{-0.2ex} \ref{Criteria A spectrum}
and \ref{Model without E38}
were determined under the minimum selection conditions (including the minimum cut level for photon
opening angles in $\gamma \gamma$ pairs, which obviously leads to systematic errors
in the low mass region due to an excess of pairs with small opening angles at the event mixing).
The results for the experimental and simulated data are shown in Fig.\hspace{-0.2ex} \ref{No cuts T0}.

% ModelWithoutE38FigA.png ModelWithoutE38FigB.png

As seen from Fig.\hspace{-0.4ex} \ref{No cuts T0}, the simulated data confirm the character of systematic errors. In addition,
an enhancement in the mass region of about 17 MeV$/c^2$ is observed in the experiment, which is absent in the model.

\begin{figure}[p]
   \centering
    \subfigure{
      \includegraphics[width=0.493\textwidth]{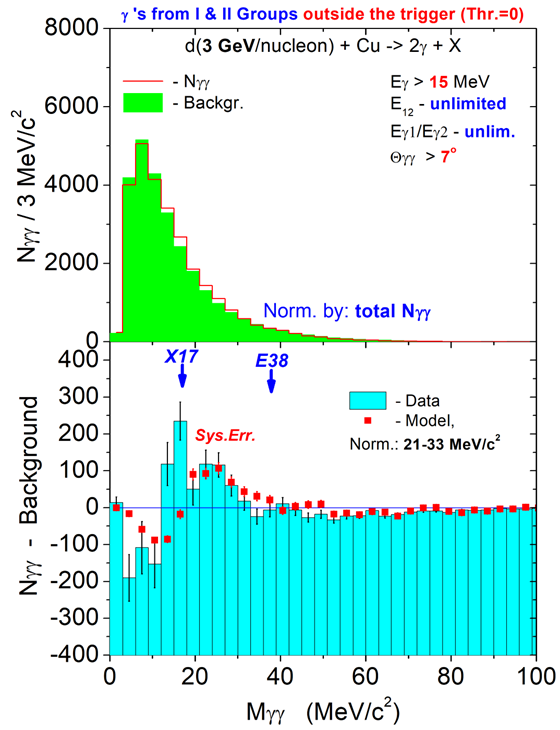}}
\hspace{-0.43cm}
    \subfigure{
      \includegraphics[width=0.497\textwidth]{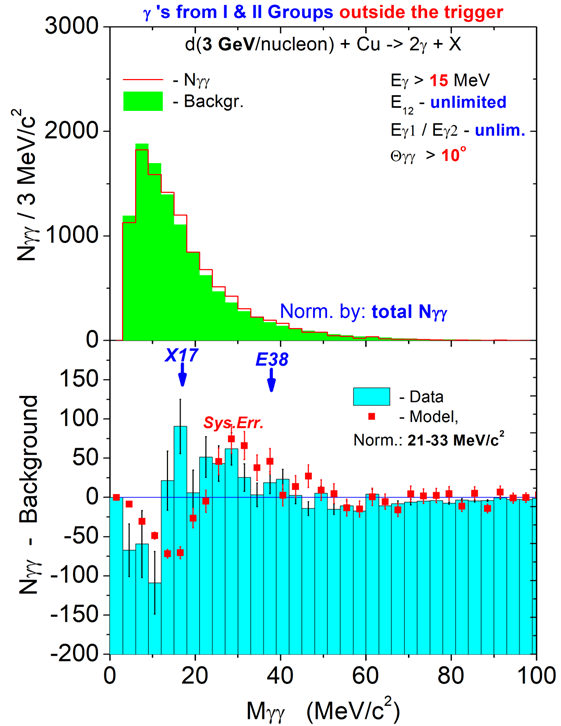}}
\caption{The same as in Fig.\hspace{-0.4ex} \ref{No cuts T0}, but after cuts by the opening angle:
$\Theta_{\gamma \gamma} > 7^{o} $ (left figure) and $\Theta_{\gamma \gamma} >
10^{o} $ (right figure).
The backgrounds are normalized to the total pair numbers in the spectra.
The spectra after the backgrounds subtraction in the experimental and simulated data
are normalized by the numbers of pairs in the range (21, 33) MeV$/c^2$.}
\label{No cuts T7T10}
\end{figure}

The same as in Fig.\hspace{-0.4ex} \ref{No cuts T0}, but after cuts by the opening angle, are shown in
Fig.\hspace{-0.2ex} \ref{No cuts T7T10}.

% ModelWithoutE38FigA.png ModelWithoutE38FigB.png

As can be seen from Fig.\hspace{-0.4ex} \ref{No cuts T7T10}, the position of the systematic errors
noticeably changes with increasing of the opening angle, while the position of the enhancement at  about 17 MeV$/c^2$ is practically stable.

\vspace{0.5cm}

\noindent{\large\bf Conclusion}

\vspace{0.2cm}

Along with $\pi^0$ mesons, signals in the form of enhanced structures at invariant masses of about 17 and 38 MeV/$c^2$ are observed in the reactions
$p + \mathrm{C}\rightarrow \gamma+\gamma+x$, $d + \mathrm{C}\rightarrow \gamma+\gamma+x$ and $d + \mathrm{Cu}\rightarrow \gamma+\gamma+x$ at momenta
5.5 GeV/c, 2.75 GeV/c and 3.83 GeV/c per nucleon, respectively.
The results of testing of the observed signals, including the results of the Monte Carlo simulation support the conclusion that the observed signals are the consequence of detection of the particles with masses of about 17 and 38 MeV/$c^2$ decaying into a pair of photons.

\begin{figure}[htbp]
\centering
\includegraphics[width=0.84\textwidth]{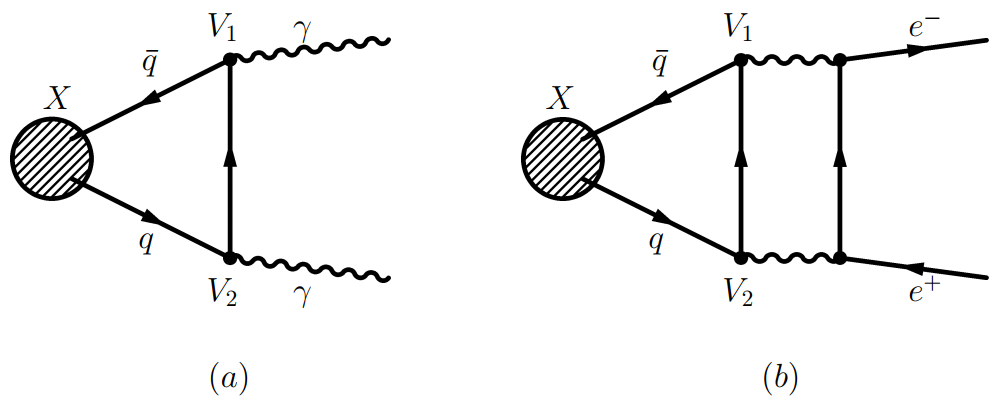}
\caption{Decay of the QED meson X into a $\gamma\gamma$ pair (a), and an $e^+e^-$ pair (b).}
\label{wongJHEP2020Figure3ab}
\end{figure}

In view of the above many theoretical possibilities, it is of great to confirm the occurrence of X17 at different initial conditions and from different decay channels.
The decay of both channels is in agreement with the composite picture of $X17$ and $E38$ \cite{Wong3Fig3}, see Fig.\ref{wongJHEP2020Figure3ab}. The our experiment corresponds to the decay Fig.\ref{wongJHEP2020Figure3ab}(a) while the ATOMKI experimental observation corresponds to the decay by Fig.\ref{wongJHEP2020Figure3ab}(b).

\begin{figure}[htbp]
\centering
\includegraphics[width=0.54\textwidth]{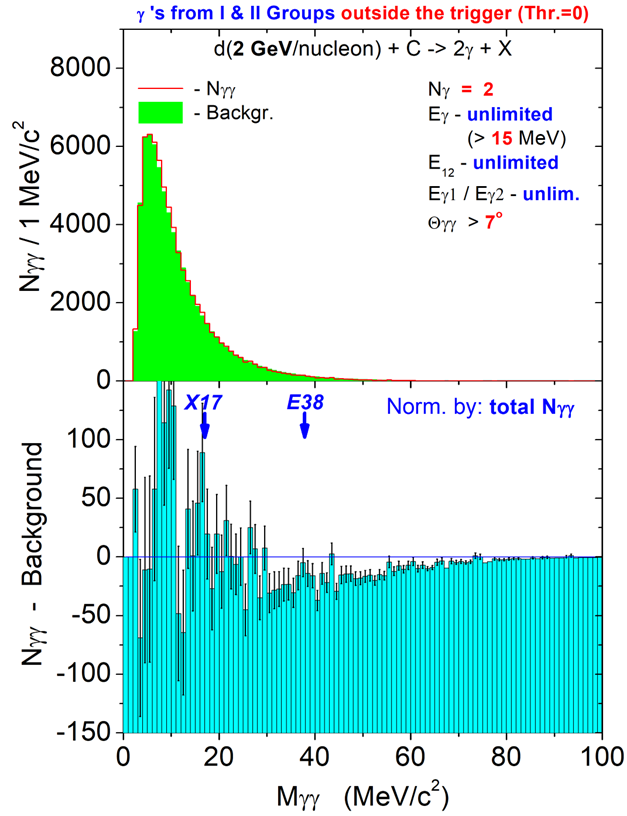}
\caption{Invariant mass distributions of $\gamma \gamma$ pairs from the reaction $d + \mathrm{C}\rightarrow \gamma+\gamma+x$ at 2.75 GeV/c per nucleon under condition $\Theta_{\gamma \gamma} > 7^{o} $ in events with the number of detected photons, $N_\gamma = 2$, without (upper panel) and with (bottom panel) the background subtraction.
The top shaded histogram shows the background contribution. The background is normalized to the total pair number
in the spectrum.}
\label{ForM9(1)}
\end{figure}

The presented evidence of both X17 and E38, together with the earlier evidence of the E38 \cite{Ab2019}, suggests that there are several particles in the anomalous region (the region of masses less than the $\pi^0$ mass). In particular, the possible existence of an anomaly at a mass of about 9 MeV/$c^2$ is also discussed \cite{ForM9}, which is beyond the scope of this article. We only point out that there is an indication of an
enhancement in the interval of 7-10 MeV/$c^2$ (see, for example, Fig.\ref{ForM9(1)}),
but the conclusion about the physical nature of the observed enhancement is preliminary, since this region can contain both noise and systematic errors, which requires further analysis.

Further, more detailed analysis of the available theoretical models and planning of new experiments are needed.

% wongJHEP2020_figure3ab.png

% ForM9(1).png

\vspace{0.3cm}

\noindent{\large\bf Acknowledgements}

\vspace{0.1cm}

We thank S. Afanasev, V. Arkhipov, A. Elishev, V. Kashirin, A. Kovalenko, A. Malakhov, and the staff of the Nuclotron
for their help in carrying out the experiments.
We are grateful to S. Gevorgyan, M. Kapishin, V. Kekelidze, A. Litvinenko, D. Madigozhin, V. Nikitin
and Yu. Potrebenikov for numerous fruitful discussions.
We are also grateful to R. Avakyan, A. Danagulyan (YSU, Yerevan), S. Barsov, O. Fedin (PNPI, Gatchina), V. Kukulin (MSU, Moscow), A. Sirunyan (YPhI, Yerevan),
O. Teryaev (LTP, JINR) for discussions and valuable remarks.
We especially thank E. van Beveren, G.Rupp, D. Blaschke and C.-Y. Wong, who initiated these studies and provided assistance in the work.

We are grateful to the organizers of the 52nd International Symposium on Multiparticle Dynamics (ISMD 2023) for the invitation and the opportunity to demonstrate our results.

The work was supported in part by the Russian Foundation for Basic Research, grants 08-02-01003-a and 11-02-01538-a.

\begin{center}
  {\large\bf Appendix 1. Analysis of amplitude spectra in separate modules}
\end{center}

\vspace{-0.3cm}

\begin{figure}[htb]
   \centering
    \subfigure{
      \includegraphics[width=0.449\textwidth]{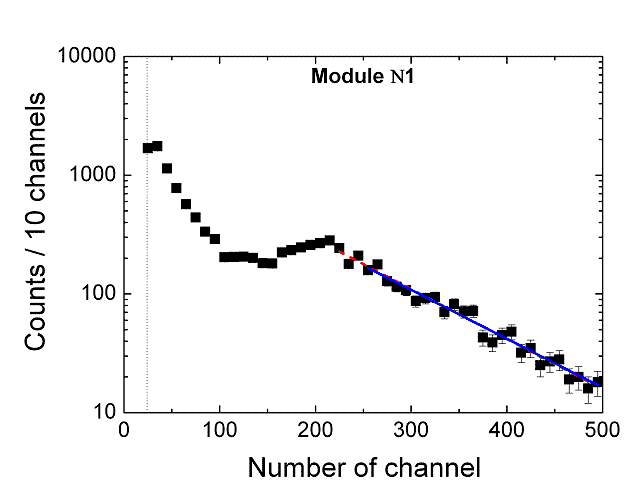}}
\hspace{-0.45cm}
    \subfigure{
      \includegraphics[width=0.449\textwidth]{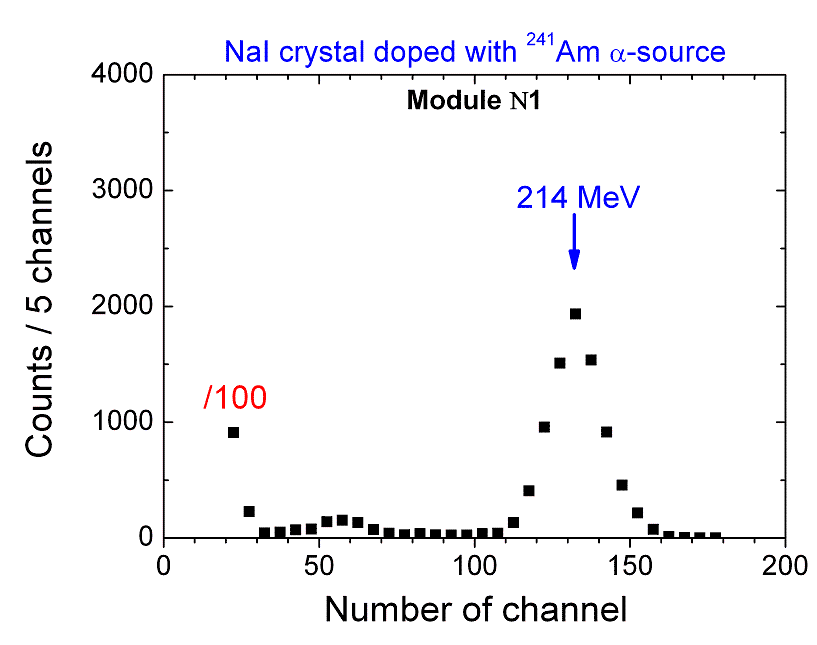}}
      \hspace{-0.45cm}
        \subfigure{
      \includegraphics[width=0.449\textwidth]{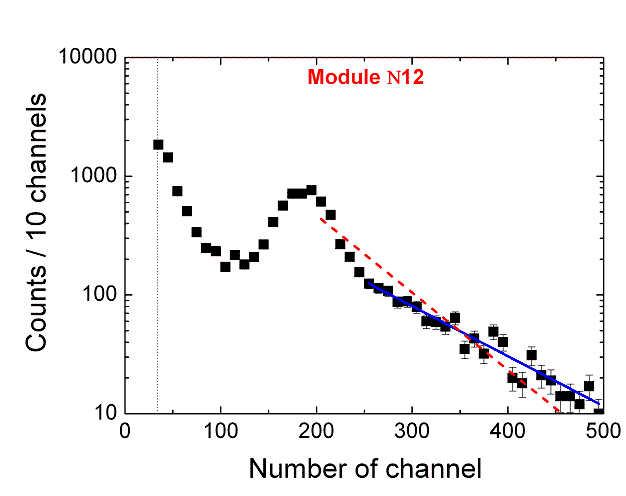}}
       \hspace{-0.45cm}
    \subfigure{
      \includegraphics[width=0.449\textwidth]{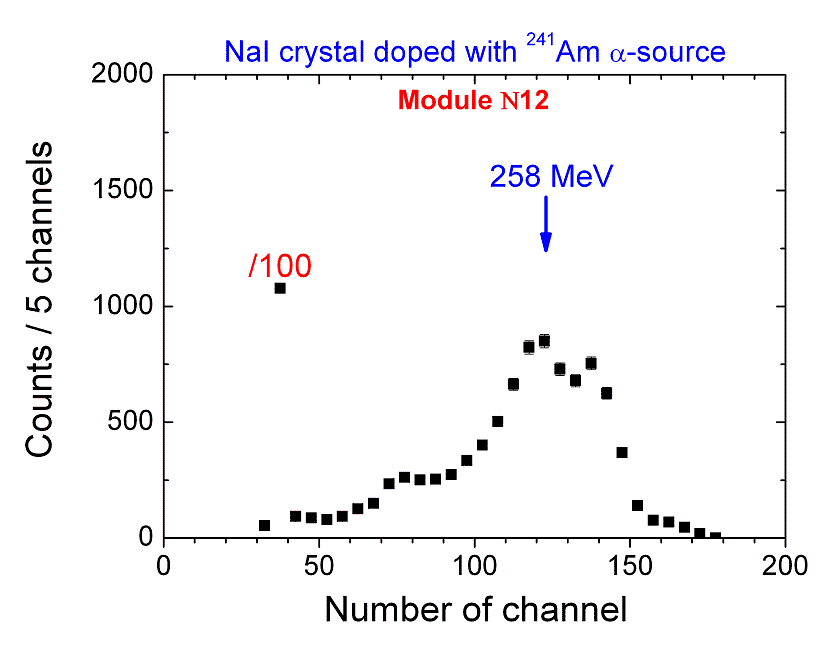}}
\caption{Amplitude spectra in separate modules in the $d + \mathrm{Cu}$ experiment (left figures) and spectra from the NaI crystals doped with $^{241}Am$ sources in the same modules (right figures) of the $\gamma$-spectrometer: in the module N1 (typical spectra as examples) and in the module N12, which was excluded from the analysis because of the poor performance. The peaks at around channel number 200  in the left figures correspond to the discriminator thresholds  (which were at the level of 0.35 GeV, see the text, section IIA.). The lines are exponential approximations in different amplitude intervals.}
\label{Modules}
\end{figure}

Typical amplitude spectra with no cuts, in separate modules in
the experiment are shown in Fig.\hspace{-0.3ex} \ref{Modules}.

\end{document}